\documentclass[12pt]{iopart}        

\usepackage{iopams}
\usepackage{graphicx}

\begin{document}

\title[]{Pyroxenes: A new class of multiferroics}

\author{S~Jodlauk$^{1}$, P~Becker$^{1}$, J~A~Mydosh$^{2}$, D~I~Khomskii$^{2}$,
T~Lorenz$^{2}$, S V Streltsov$^{2,4}$, D C Hezel$^{3}$ and L
Bohat\'{y}$^{1,5}$\footnote[0]{$^4$ Permanent address: Inst.\ of
Metal Phys., S. Kovalevskoy Street 18,
620219 Ekaterinburg GSP-170, Russia. \\
$^5$ Author to whom correspondence should be addressed
(ladislav.bohaty@uni-koeln.de).} }

\address{$^1$ Institut f\"{u}r Kristallographie,
Universit\"{a}t zu K\"{o}ln, Z\"{u}lpicher Str.\ 49 b, 50674 K\"{o}ln, Germany.}

\address{$^2$ II.\,Physikalisches Institut, Universit\"{a}t zu K\"{o}ln, Z\"{u}lpicher
Str.\ 77, 50937 K\"{o}ln, Germany.}

\address{$^3$ Institut f\"{u}r Geologie und Mineralogie,
Universit\"{a}t zu K\"{o}ln, Z\"{u}lpicher Str.\ 49 b, 50674 K\"{o}ln, Germany.}

\begin{abstract}
Pyroxenes with the general formula $AM$Si$_2$O$_6$ ($A$ = mono- or
divalent metal, $M$ = di- or trivalent metal) are shown to be a
new class of multiferroic materials. In particular, we have found
so far that NaFeSi$_2$O$_6$ becomes ferroelectric in a
magnetically ordered state below $\simeq 6$~K. Similarly,
magnetically driven ferroelectricity is also detected in the Li
homologues, LiFeSi$_2$O$_6$ ($T_C \simeq 18$~K) and
LiCrSi$_2$O$_6$ ($T_C \simeq 11$~K). In all these monoclinic
systems the electric polarization can be strongly modified by
magnetic fields. Measurements of magnetic susceptibility,
pyroelectric current and dielectric constants (and their
dependence on magnetic field) are performed using a natural
crystal of aegirine (NaFeSi$_2$O$_6$) and synthetic crystals of
LiFeSi$_2$O$_6$ and LiCrSi$_2$O$_6$ grown from melt solution. For
NaFeSi$_2$O$_6$ a temperature versus magnetic field phase diagram
for NaFeSi$_2$O$_6$ is proposed. Exchange constants are computed
on the basis of {\it ab initio} band structure calculations. The
possibility of a spiral magnetic structure caused by frustration
as origin of the multiferroic behaviour is discussed. We propose
that other pyroxenes may also be multiferroic, and that the
versatility of this family offers an exceptional opportunity to
study general conditions for and mechanisms of magnetically
driven ferroelectricity.
\end{abstract}




\section{Introduction}

Multiferroic materials, which are simultaneously (ferro)magnetic,
ferroelectric and  ferroelastic, have very interesting physical
properties and promise important applications. They are presently
attracting considerable attention [1-6]. There exist several
different classes of multiferroics [5], a very interesting type
being the recently discovered [7, 8] systems in which
ferroelectricity appears only in certain magnetically ordered
states, typically, although not necessarily,  spiral ones [6].
Even though their electric polarization $\mathbf{P}$ is usually
not large, one can easily influence it by comparatively weak
magnetic fields. It is this magnetic "switching" of $\mathbf{P}$
which makes multiferroics potentially very useful in device
applications. At present relatively few such materials are known:
some of them [7-11] are multiferroic in zero magnetic field,
while others develop an electric polarization only if a magnetic
field is applied [12-14]. These compounds belong to different
crystallographic classes, and although some general rules
governing their behaviour are already established [6], there is
as yet no complete or general understanding of the origin of
multiferroic behaviour.

Here we report the discovery and study of a new class of
multiferroics, which opens a fresh possibility to investigate the
systematics of this phenomenon. Interestingly, this class --
pyroxenes -- form a very important group of minerals: They
provide more than 20 vol.\% of the Earth's crust and upper mantle
to a depth of 400 km [15, 16]. In addition, with certain
combinations of cations they create such well-known semiprecious
stones as the famous Chinese jade and they even are found as
constituents of extraterrestrial materials such as lunar and
Martian rocks and meteorites [17, 18]. Pyroxenes belong to the
silicates of the general composition $AM$Si$_2$O$_6$  where $A$
stands for mono- or divalent metals while $M$ represents di- or
trivalent metals. Their crystal structures possess orthorhombic or
monoclinic symmetries which can accept a wide variety of $M$ and
$A$ elements, especially the $3d$-transition metals. In most cases
Si may be also substituted by Ge. Thus, pyroxenes offer a quite
broad and flexible class of materials for physical
investigations. Until now detailed investigations of these
systems have been mainly focused on mineralogical and
crystallographic aspects. Only recently did their magnetic
properties attract some attention, in particular due to the
observation of the orbitally-driven spin gap state in
NaTiSi$_2$O$_6$ [19, 20]. Most pyroxenes seem to be
antiferromagnetic [21], yet complete magnetic structures are only
known for few members of this class.

The specific features and the variety of magnetic properties of
pyroxenes are determined by their crystal structure, shown in
figure 1. The main building blocks are one-dimensional zig-zag
chains of edge-sharing [$M$O$_6$] octahedra running along the
crystallographic $c$-axis. Within the (110) and ($\overline{1}10$)
planes these chains are connected by chains of [SiO$_4$] (or
[GeO$_4$]) tetrahedra. Important here is that, besides the
quasi-one-dimensionality of the [$M$O$_6$] chains, the relative
packing of the magnetic chains forms a triangular-type magnetic
lattice in each (110) plane, see figure 1c, giving rise to a
magnetic frustration. A similar frustration is present in the
($\overline{1}10$) plane. Such geometric frustration can in
principle lead to complex magnetic structures, in particular,
commensurate or incommensurate spiral ordering, which, according
to our present understanding, may be favourable for
magnetically-induced ferroelectricity [6]. So based upon their
crystal structure pyroxenes are excellent candidates for
multiferroic behaviour. Another beneficial feature of these
materials is that they are usually good insulators (transparent
crystals of green, yellow or brown colour), which is important
for ferroelectric materials.

We find that indeed at least three members of this family,
NaFeSi$_2$O$_6$, LiFeSi$_2$O$_6$ and LiCrSi$_2$O$_6$ become
ferroelectric in a magnetically ordered state. In all of them the
ferroelectricity can be strongly modified by the application of
magnetic fields along different crystallographic directions.

\begin{figure}[t]
\hfill
\includegraphics[width=0.83\textwidth]{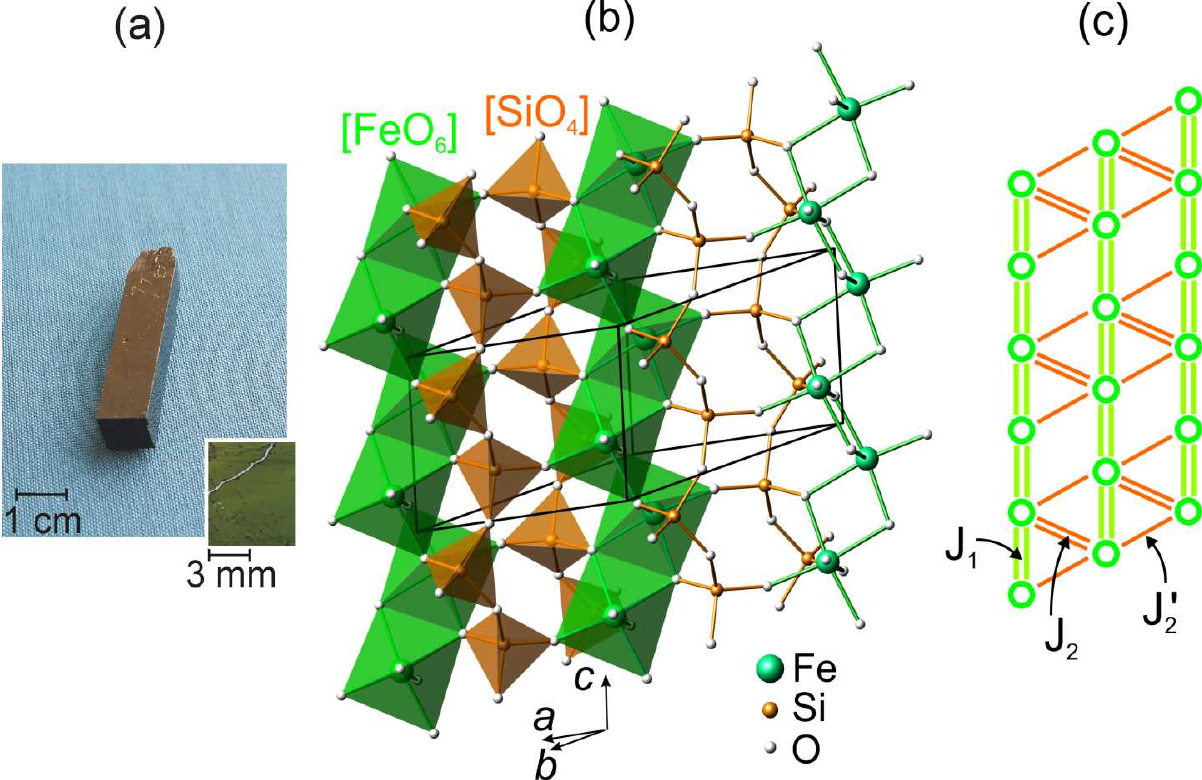}
\caption{\label{fig1} \textbf{(a)} Natural aegirine
(NaFeSi$_2$O$_6$) crystal from a pegmatite of alkaline rocks of
Mount Malosa, Malawi. In thin sections (see inset) the aegirine
crystal is green and transparent. \textbf{(b)} The main features
of the crystal structure are chains of edge-sharing [FeO$_6$]
octahedra (green) and chains of corner-sharing [SiO$_4$]
tetrahedra (orange) running along the $c$-direction (structure
data from [22]). The linkage between iron atoms of neighboring
chains is visualized in the right part using the stick-and-ball
representation. Sodium atoms are left out for clarity. Black
lines denote the unit cell. The crystal structures of
LiFeSi$_2$O$_6$ and LiCrSi$_2$O$_6$ are similar, but with space
group symmetry $P2_1/c$ [23], while it is $C2/c$ for
NaFeSi$_2$O$_6$. \textbf{(c)} Schematic elements of the magnetic
subsystem. The double green lines represent the intrachain
exchange interaction $J_1$ arising from two equivalent Fe--O--Fe
paths. Despite the geometric zig-zag arrangement of the [FeO$_6$]
octahedra, one may expect for Fe$^{3+}$ a uniform $J_1$, since
both the bond distances and the bond angles do not alternate
along the chain direction and all $3d$-orbitals of Fe$^{3+}$ are
half-filled. The interchain exchange $J_2$ (double yellow lines)
and $J'_2$ (single yellow lines) arise from the Fe--Fe coupling
via two and one [SiO$_4$] tetrahedra, respectively.}
\end{figure}

\section{Experimental details and methods}

NaFeSi$_2$O$_6$ has been studied on different samples cut from a
high-quality natural single crystal from Mount Malosa, Malawi, of
size $ 10 \times 10 \times 80$~mm$^3$, while for LiFeSi$_2$O$_6$
and LiCrSi$_2$O$_6$ we used synthetic single crystals. The
chemical composition and homogeneity of our natural crystal of
aegirine were analyzed by electron microprobe (JEOL 8900RL) on a
(001) plate yielding the compositional formula
Na$_{1.04}$Fe$_{0.83}$Ca$_{0.04}$Mn$_{0.02}$Al$_{0.01}$Ti$_{0.08}$Si$_{2}$O$_{6}$.
Transparent single crystals of LiFeSi$_2$O$_6$ (light green) and
LiCrSi$_2$O$_6$ (emerald-green) were grown from melt solution with
dimensions reaching $4 \times 2 \times 0.5$~mm$^3$.

The magnetic susceptibility, pyroelectric current (typical peak
heights: 50~fA--5~pA, typical baselines: 20--400~fA with noise of
order 10~fA) and capacitance were measured by a vibrating sample
magnetometer (Quantum Design PPMS), an electrometer (Keithley
6517A) and a capacitance bridge (Andeen-Hagerling 2500A),
respectively. For the dielectric investigations plate-like
samples coated with gold or silver electrodes were used. The
dielectric constant $\varepsilon^r$ was calculated from the
measured capacitance. The pyroelectric current was recorded
without applied electric field at a heating rate of $+1$~K/min
for NaFeSi$_2$O$_6$ and $+4$~K/min for LiFeSi$_2$O$_6$ and
LiCrSi$_2$O$_6$ after having cooled the sample in a static
electric poling field of at least 200~V/mm. Typical leakage
currents caused by the poling field amount to about 50~fA,
meaning that the resistivities of our crystals are larger than
$10^{14}~\Omega$m. By reversing the poling field the pyroelectric
origin of the measured current and the ferroelectric switching of
polarization was established. The polarization was completely
reversible in NaFeSi$_2$O$_6$ and in LiFeSi$_2$O$_6$. In our small
single crystals of LiCrSi$_2$O$_6$ the full reversal of
pyrocurrent was not reached but only a reduction by $\simeq
50$\%, most probably due to the domain wall pinning. This feature
has to be checked on larger crystals. For calculating the
polarization $\mathbf{P}$ by integration of the pyroelectric
current, a non-polarization-caused baseline (dependent on sample
and measurement geometry) was eliminated.

For the band structure calculations we used the tight-binding
linearized muffin-tin (MT) orbitals method and the LSDA+U
approximation, which takes into account the on-site Coulomb
correlations ($U$) in a mean-field way. We utilized the values of
the Hubbard $U = 4.5$~eV and Hund's rule $J_H = 1$~eV obtained by
the same calculation scheme described in [24]. The exchange
interaction parameters were computed as a second derivative of
the energy variation at small spin rotations [25].

\begin{figure}[t]
\hfill
\includegraphics[width=\textwidth]{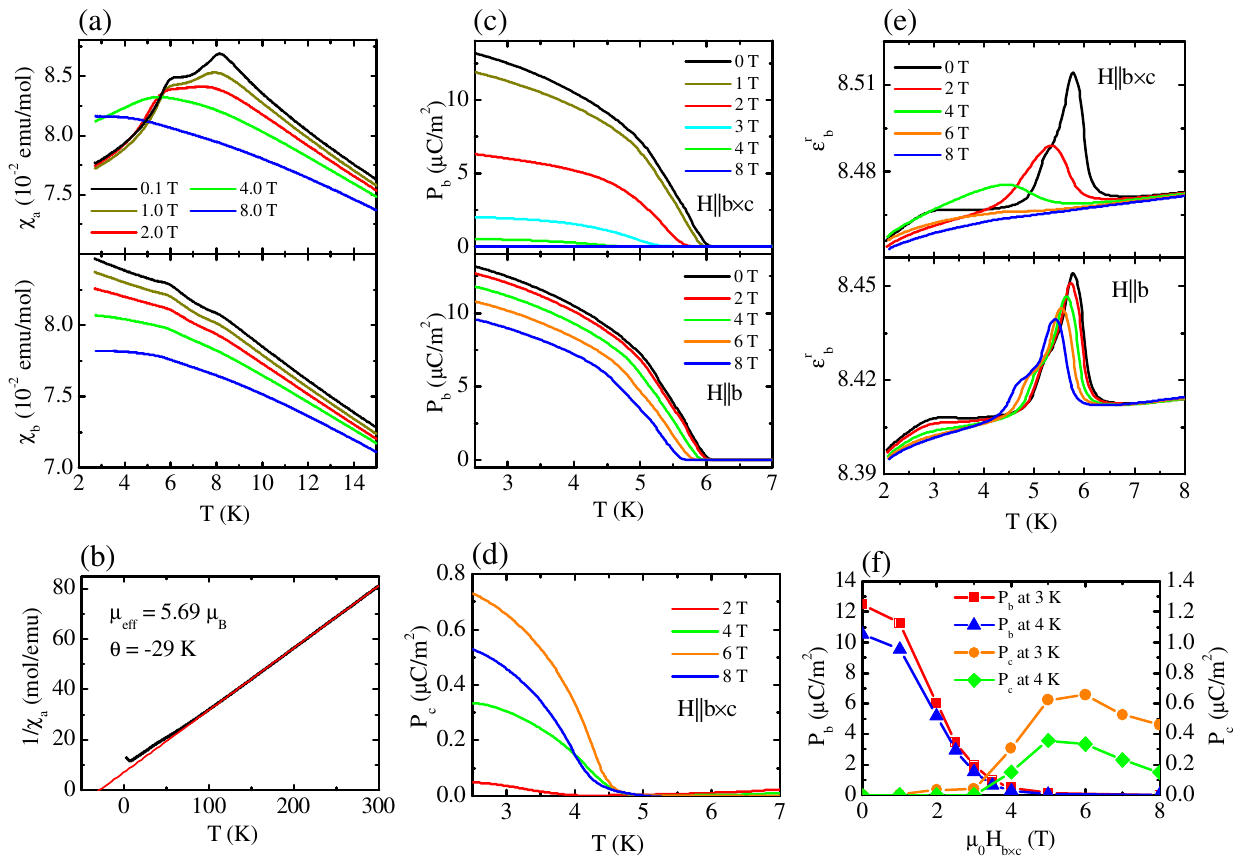}
\caption{\label{fig2} Magnetic and ferroelectric properties of
NaFeSi$_2$O$_6$: Representative temperature dependences of the
magnetic susceptibility (\textbf{(a)} and \textbf{(b)}),
spontaneous electric polarization (\textbf{(c)} and \textbf{(d)})
and the longitudinal component of the dielectric tensor
$\varepsilon_b^r$ \textbf{(e)} measured for different magnetic
field strengths and orientations as indicated in the figure.
Panel \textbf{(f)} displays the polarizations $\mathbf{P_b}$ and
$\mathbf{P_c}$ as a function of a magnetic field $\mathbf{H}||
(\mathbf{b}\times \mathbf{c})$ at selected temperatures.}
\end{figure}

\section{Results}
\subsection{NaFeSi$_2$O$_6$}

On NaFeSi$_2$O$_6$ (that is monoclinic with space group symmetry
$C2/c$ in the paramagnetic state [22]) we measured the magnetic
susceptibility in magnetic fields applied along all three
crystallographic axes. The results for $\chi_a$ and $\chi_b$ are
shown in figure 2a. The measurements of $\chi_c$ resemble those
for $\chi_a$ and are not shown. The inverse susceptibility
$1/\chi_a$ taken in a field of 1~Tesla is presented in figure 2b.
The susceptibility data show nearly paramagnetic (Curie-Weiss)
behaviour down to a temperature of about 50 K and the clear onset
of antiferromagnetic ordering at 8 K. A negative Weiss
temperature $\Theta\simeq -29$~K signals a net antiferromagnetic
exchange interaction. The value of the effective magnetic moment
$\mu_{eff}$ amounts to 5.69\,$\mu_B$ close to the theoretically
expected value of $g\sqrt{S(S+1)}\mu_B = 5.92$\,$\mu_B$ for the
spin-only value of the Fe$^{3+}$ ($S=5/2$) magnetic moments. While
$\chi_a$ (and $\chi_c$) for low magnetic fields exhibit a maximum
indicating the antiferromagnetic order, only a kink is observed
for $\chi_b$. This is a strong indication, that (at least for low
magnetic fields) the spins of the Fe$^{3+}$-ions lie within the
$ac$-plane. By applying higher magnetic fields the transition
temperature ($T_N$) is reduced and the anomaly strongly broadens
until it is not resolvable anymore in a field of 8 Tesla. Another
intriguing feature of the susceptibility data is an additional
anomaly around 6~K. Its position slightly decreases towards lower
temperatures with increasing field and it vanishes for fields of
more than about 4 Tesla. As will be illustrated below, this
behaviour is related to the onset of a spontaneous electrical
polarization in NaFeSi$_2$O$_6$.

\begin{figure}[t]
\begin{center}
\includegraphics[width=0.5\textwidth]{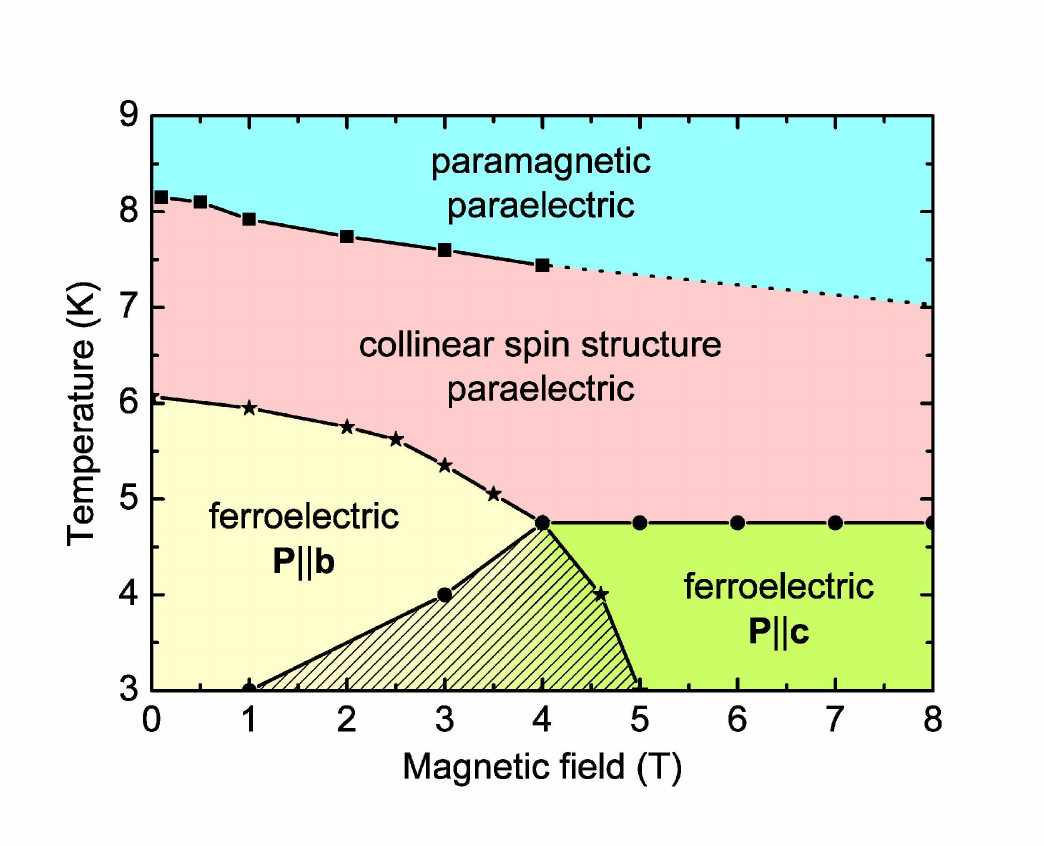}
\end{center}
\caption{\label{fig3} Temperature versus magnetic field (applied
in the $ac$-plane) phase diagram for NaFeSi$_2$O$_6$ illustrating
the multiferroic behaviour.}
\end{figure}

Measurements of the pyroelectric current and of the dielectric
constants were performed on three mutually perpendicular
plate-like samples (sample surfaces range from 40 -- 80~mm$^2$,
typical thickness is about 1~mm) with face normals along
$\mathbf{b}$, $\mathbf{c}$, and $(\mathbf{b}\times \mathbf{c})$,
which does not coincide with $\mathbf{a}$ in a monoclinic unit
cell. For each sample the magnetic field was applied along
$\mathbf{b}$, $\mathbf{c}$, and $(\mathbf{b}\times \mathbf{c})$.
In figures 2c and 2d we present the electric polarization
$\mathbf{P_b}$ along $\mathbf{b}$ and $\mathbf{P_c}$ along
$\mathbf{c}$ measured in different magnetic fields $\mathbf{H}$
applied either along  the $(\mathbf{b}\times \mathbf{c})$ or the
$\mathbf{b}$ direction ($\mathbf{P_b}$ for $\mathbf{H} ||
\mathbf{c}$ is similar to $\mathbf{P_b}$ for $\mathbf{H} ||
(\mathbf{b}\times \mathbf{c})$ and not shown).  From these data
we recognize that NaFeSi$_2$O$_6$ becomes ferroelectric below
$T_{FE} = 6$~K, with the polarization $\mathbf{P} || \mathbf{b}$.
The onset temperature $T_{FE}$ and the magnitude of $\mathbf{P_b}$
are suppressed by the field  $\mathbf{H} || (\mathbf{b}\times
\mathbf{c})$, but hardly change for $\mathbf{H} || \mathbf{b}$.
This is confirmed by a well-defined peak of the longitudinal
component $\varepsilon_b^r$ of the relative dielectric tensor
along $\mathbf{b}$ (figure 2e), which is strongly suppressed by a
magnetic field applied within the $ac$-plane, but only weakly
shifts to lower temperature for $\mathbf{H} || \mathbf{b}$. This
anisotropic magnetic-field dependence of the ferroelectric
ordering fully corresponds to that of the magnetic ordering
observed in the susceptibility measurements. When the
polarization $\mathbf{P_b}$ is suppressed by $\mathbf{H}
||(\mathbf{b}\times \mathbf{c})$, a (smaller) spontaneous
polarization $\mathbf{P_c}$ appears instead. A similar
polarization $\mathbf{P_c}$ is also generated by $\mathbf{H} ||
\mathbf{c}$. Thus, a field $\mathbf{H}$ applied within the
$ac$-plane leads to a gradual rotation of the polarization from
$\mathbf{b}$ to $\mathbf{c}$, see figure 2f, while a magnetic
field applied along the $b$-direction does not create any
measurable $\mathbf{P_c}$. An electric polarization
$\mathbf{P}_{\mathbf{b}\times \mathbf{c}}$ along
$(\mathbf{b}\times \mathbf{c})$ could not be detected,
independent of the direction and strength of the applied magnetic
field. The proposed phase diagram of the magnetoelectric
behaviour of NaFeSi$_2$O$_6$ is given in figure 3.

\subsection{LiFeSi$_2$O$_6$ and LiCrSi$_2$O$_6$}

\begin{figure}[t]
\hfill
\includegraphics[width=0.85\textwidth]{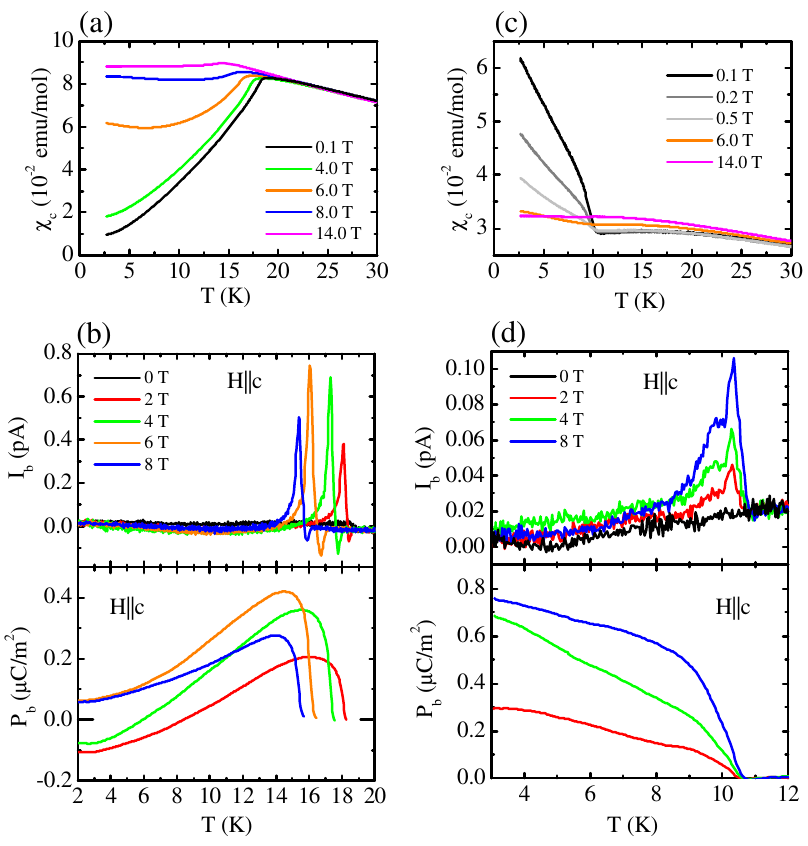}
\caption{\label{fig4} Selected temperature dependences of the
magnetic susceptibility $\chi_c$, pyroelectric current
$\mathbf{I_b}$ and electric polarization $\mathbf{P_b}$ of
LiFeSi$_2$O$_6$ (\textbf{(a)} and \textbf{(b)}) and
LiCrSi$_2$O$_6$ (\textbf{(c)} and \textbf{(d)}) for different
magnetic fields applied along the $c$-direction. The small
current peaks of negative sign in panel \textbf{(b)} are not of
pyroelectric origin (see section 2). }
\end{figure}

For the Li-pyroxenes LiFeSi$_2$O$_6$ and LiCrSi$_2$O$_6$ (with
space group symmetry $P2_1/c$ below 228~K and 335~K,
respectively, and $C2/c$ above these temperatures [26]) we
measured $\chi_c$ as well as the pyroelectric current and
dielectric constants on thin ($\simeq 0.5$~mm) plate-like samples
with \{010\} surfaces of about 10~mm$^2$ and 4~mm$^2$,
respectively, in magnetic fields applied along $\mathbf{c}$. As
presented in figure 4a, LiFeSi$_2$O$_6$ shows paramagnetic
behaviour down to a temperature of 18~K, where at low applied
magnetic field the clear onset of antiferromagnetic ordering is
observed. The onset temperature gradually decreases to 14 K when
increasing the magnetic field to 14~Tesla. The strong
low-temperature increase of $\chi_c$ around 6~Tesla indicates a
spin-flop transition, where the orientation of the spins changes
from parallel $\mathbf{c}$ to perpendicular $\mathbf{c}$.
Figure~4b presents the measured pyroelectric current and the
corresponding electric polarization along the $b$-direction of the
crystal. While there is essentially no pyroelectric current in
zero magnetic field, we find pronounced peaks in the pyroelectric
current for finite fields. For $H > 0$, we also observe small
precursors of negative sign in the pyroelectric current. Their
origin is not yet clear, but they cannot be attributed to
ferroelectric behaviour, since -- in contrast to the large peaks
-- they neither change sign nor shape or size when the electric
poling field is reversed. With increasing magnetic field these
peaks shift to lower temperatures similar to the shift of the
antiferromagnetic ordering temperatures observed in $\chi_c$.
Integrating the difference of the pyroelectric current in finite
and zero magnetic field reveals a clear development of a
spontaneous electric polarization $\mathbf{P_b}$ below the
magnetic ordering temperature in finite magnetic fields. However,
due to small negative pyroelectric currents in finite fields for
temperatures below the peak, the calculated polarization
decreases again and even changes sign with further decreasing
temperature. Whether or not this unexpected behaviour is
intrinsic requires further studies on larger crystals.

The magnetic susceptibility $\chi_c$ of LiCrSi$_2$O$_6$ (figure
4c) shows a different behaviour. Around 20~K we observe a broad
maximum of $\chi_c$, as is typical for low-dimensional spin
systems. Below 11~K, a sharp increase of $\chi_c$ sets in
indicating a long-range magnetic order with a small ferromagnetic
component. This indicates a slightly canted antiferromagnetic spin
structure. The maximum spontaneous magnetic moment of about
$0.005\,\mu_B$ suggests a canting angle of about $0.1^\circ$ of
the total spin moment of $3\,\mu_B$ per Cr$^{3+}$. Similar to
LiFeSi$_2$O$_6$, we find an electric polarization $\mathbf{P_b}$
along $\mathbf{b}$ in an applied magnetic field along $\mathbf{c}$
while $\mathbf{P_b}$ remains zero without a magnetic field. The
pyroelectric current and the resulting electric polarization of
LiCrSi$_2$O$_6$ (figure 4d) show a sharp phase transition
slightly below 11~K that coincides with the temperature of
magnetic ordering. In contrast to LiFeSi$_2$O$_6$ we observe
neither a shift of the onset temperatures of magnetic ordering
nor of the electric polarization with increasing magnetic field.
The magnitude of the electric polarization monotonically grows
with increasing magnetic field up to the highest applied field of
8~Tesla.

\section{Discussion}

According to the data presented above we identify at least three
members of the large pyroxene family that develop
magnetically-driven ferroelectricity, which strongly depends on
the applied magnetic field. The ferroelectric behaviour was
established by reversing the static electric poling field (see
section 2). The explanation of the appearance of ferroelectricity
in magnetic states requires the specific knowledge of the
corresponding magnetic structure. The early data on the magnetic
structure of NaFeSi$_2$O$_6$ are presented in [22, 27], and those
for LiFeSi$_2$O$_6$ in [27, 28] (for LiCrSi$_2$O$_6$ no
information on the magnetic structure is available in
literature). The proposed structures, all being
antiferromagnetic, are nevertheless different, even for the same
material. In [22] it was suggested that in aegirine the spins of
the one-dimensional zig-zag chains in the $c$-direction are
ferromagnetic, with the neighbouring chains being antiparallel.
In [27] the same magnetic coupling scheme, but also a model of
purely antiferromagnetic coupling within and between the chains,
was considered for both, NaFeSi$_2$O$_6$ and LiFeSi$_2$O$_6$,
based upon magnetization and M\"{o}ssbauer investigations. And in
[28] the authors conclude that the spins are
antiferromagnetically ordered both intrachain and interchain for
LiFeSi$_2$O$_6$, but the detailed magnetic structure was not
revealed. At the same time several peaks in the neutron
scattering spectra of NaFeSi$_2$O$_6$ observed in [22] remain
unexplained, and the authors themselves speculate that they may
be connected with a possible incommensurate superstructure caused
by frustration. In the absence of unambiguous data on the
magnetic structure we can only assume that there may indeed exist
a spiral structure in these materials, possibly in addition to
the main features found in [22, 27, 28]. Actually, a spiral
structure should be expected from the "triangular" topology seen
in figure 1c and also from the values of the exchange constants.
The latter were determined from {\it ab-initio} band structure
calculations which we carried out using the LSDA+U method (see
section~2). For NaFeSi$_2$O$_6$ we obtained the intrachain
exchange integral $J_1 = 8.5$~K and the interchain coupling $J_2
= 1.6$~K; $J'_2= 0.8$~K, and for LiFeSi$_2$O$_6$ $J_1 = 7.0$~K,
$J_2 = 3.4$~K;  $J'_2= 1.9$~K --- {\it all antiferromagnetic}, in
full agreement with the Goodenough-Kanamori rules. For such
values of exchange one expects a spiral magnetic structure for the
triangular lattice [29].

Based upon our exchange scheme we can propose a scenario which
would explain the appearance of ferroelectricity by the existence
of spiral magnetic structures in NaFeSi$_2$O$_6$ (and probably in
LiFeSi$_2$O$_6$ and LiCrSi$_2$O$_6$ as well), consistent with the
general considerations of [6]. However, we can not exclude other
possible mechanisms, e.g.\ magnetostriction [5, 6]. The features
of the observed polarization in all three materials,
NaFeSi$_2$O$_6$, LiFeSi$_2$O$_6$ and LiCrSi$_2$O$_6$, indicate
that the crystals remain monoclinic in their magnetically and
electrically ordered phase, and therefore the phase transition
follows the simplest group-subgroup path. In NaFeSi$_2$O$_6$ the
loss of inversion symmetry of the paraelectric phase (point group
$2/m$) at $T_{FE}$ results in the polar point groups $2$
($\mathbf{P} || \mathbf{b}$) or $m$ ($\mathbf{P}$ perpendicular
$\mathbf{b}$), depending upon the applied magnetic field. So far
in the Li-compounds spontaneous polarization was observed
parallel $\mathbf{b}$ in finite field, but it is not clear
whether for $H = 0$ the polarization is simply absent (or very
small), or along another direction. At present our available
single crystals do not allow a definite conclusion. Another open
question is whether the presence of non-centrosymmetric [SiO$_4$]
tetrahedra plays some role in providing the mechanism of
ferroelectricity in pyroxenes.

\section{Conclusions}

We have found {\it a new class of multiferroic materials} among
the geologically important pyroxenes with the general formula
$AM$Si$_2$O$_6$. So far, three members of this large family of
compounds --- NaFeSi$_2$O$_6$, LiFeSi$_2$O$_6$ and LiCrSi$_2$O$_6$
--- show magnetically-induced ferroelectricity. Both the magnitude
and direction of the polarization can be strongly modified by a
magnetic field. In the absence of reliable data on magnetic
structure we can not exactly identify the source of multiferroic
behaviour in the pyroxenes, but we suggest that its origin may be
connected with a spiral magnetic structure caused by their
magnetic frustration. The existence of many compounds with the
pyroxene structure advances the hope that other materials of this
large family would also display multiferroic behaviour at yet
higher temperatures. The fact that this phenomenon is observed in
materials that are very important in geology adds a new
ingredient to our findings. One can only speculate whether the
physics disclosed here may have important outgrowths for
geophysics, e.g.\ concerning cold extraterrestrial objects.
However, even irrespective of this potentially important
interplay, the very fact of the observation of a whole new class
of (nearly) isostructural multiferroic materials has a
substantial interest and should lead to a new and better
understanding of this fascinating phenomenon.

\ack

This work was supported by the Deutsche Forschungsgemeinschaft
via the Sonderforschungsbereich 608.


\section*{References}


\begin{thebibliography}{99}

\bibitem{P1} Fiebig M 2005 \textit{J. Phys. }D \textbf{38} R123-52

\bibitem{P2} Tokura Y 2006 \textit{Science} \textbf{312} 1481-2

\bibitem{P3} Eerenstein W, Mazur N D and Scott J F
2006 \textit{Nature} \textbf{442} 759-65

\bibitem{P4}     Spaldin N A and Fiebig M
2006 \textbf{Science} \textbf{309} 391-2

\bibitem{P5} Khomskii D I 2006 \textit{J. Magn.
Magn. Mater.} \textbf{306} 1-8

\bibitem{P6} Cheong S-W and Mostovoy M V
2007 \textit{Nature Materials} \textbf{6} 13-20

\bibitem{P7}     Kimura T, Goto T,
Shintani H, Ishizaka K, Arima T and Tokura Y 2003 \textit{Nature}
\textbf{426}, 55-8

\bibitem{P8} Hur N, Park S, Sharma P A, Ahn J S, Guha S and
Cheong S-W 2004 \textit{Nature} \textbf{429} 392-5

\bibitem{P9} Lawes G et al. 2005
\textit{Phys. Rev. Lett.} \textbf{95} 087205

\bibitem{P10} Heyer O, Hollmann N,
Klassen I, Jodlauk S, Bohat\'{y} L, Becker P, Mydosh J A, Lorenz T
and Khomskii D 2006 \textit{J. Phys.: Condens. Matter}
\textbf{18} L471-5

\bibitem{P11} Park S, Choi Y J, Zhang C L and Cheong S-W 2007
\textit{Phys. Rev. Lett.} \textbf{98} 057601

\bibitem{P12} Kimura T, Lawes G and
Ramirez A P 2005 \textit{Phys. Rev. Lett.} \textbf{94} 137201

\bibitem{P13} Baier
J, Meier D, Berggold K, Hemberger J, Balbashov A, Mydosh J A and
Lorenz T 2006 \textit{Phys. Rev.} B \textbf{73} 100402(R)

\bibitem{P14} Kimura T,
Lashley J C and Ramirez A P 2006 \textit{Phys. Rev.} B \textbf{73}
220401(R)

\bibitem{P15} Deer W A, Howie R A and Zussman, J 2001 \textit{Single Chain
Silicates} (\textit{Rock-forming minerals}, vol. 2A) (London:
Longman)

\bibitem{P16} Ringwood A E 1991 \textit{Geochim Cosmochim. Acta} \textbf{55}
2083-110

\bibitem{P17} Papike J J (ed) 1999 \textit{Planetary materials}
(\textit{Reviews in Mineralogy} vol. 36) (Washington DC:
Mineralogical Society of America)

\bibitem{P18}    Jolliff B L, Wieczorek M A,
Shearer Ch K and Neal C R (ed) 2006 \textit{New views of the moon}
(\textit{Reviews in Mineralogy \& Geochemistry} vol. 60)
(Washington DC: Mineralogical Society of America)

\bibitem{P19}    Isobe M,
Ninomiya E, Vasil'ev A N and Ueda Y 2002 \textit{J. Phys. Soc.
Japan} \textbf{71} 1423-6

\bibitem{P20}    Streltsov S V, Popova O A and Khomskii D
I 2006 \textit{Phys. Rev. Lett.} \textbf{96} 249701

\bibitem{P21} Vasiliev A N,
Ignatchik O L, Sokolov A N, Hiroi Z, Isobe M and Ueda Y 2005
\textit{Phys. Rev.} B \textbf{72} 012412

\bibitem{P22}    Ballet O, Coey J M D,
Fillion G, Ghose A, Hewat A and Regnard J R 1989 \textit{Phys.
Chem. Minerals} \textbf{16} 672-7

\bibitem{P23} Redhammer G J and Roth G 2004 \textit{Z.
Kristallogr.} \textbf{219} 278-94

\bibitem{P24} Leonov I, Yaresko A N,
Antonov V N, Korotin M A and Anisimov V I 2004 \textit{Phys. Rev.
Lett.} \textbf{93} 146404

\bibitem{P25}    Katsnelson M I and Lichtenstein A I
2000 \textit{Phys. Rev.} B \textbf{61} 8906

\bibitem{P26}    Behruzi M, Hahn T,
Prewitt C T and Baldwin K 1984 \textit{Acta Crystallogr.} A
\textbf{40} \textit{Suppl.} C-247

\bibitem{P27} Baum E, Treutmann W, Behruzi M, Lottermoser W
and Amthauer G 1988 \textit{Z. Kristallogr.} \textbf{183} 273-84

\bibitem{P28}
Redhammer G J, Roth G, Paulus W, Andr\'{e} G, Lottermoser W, Amthauer
G, Treutmann W and Koppelhuber-Bitschnau B 2001 P\textit{hys.
Chem. Minerals} \textbf{28} 337-46

\bibitem{P29}    Zhang W, Saslow W M and Gabay
M 1991 \textit{Phys. Rev.} B \textbf{44} 5129-31



\end{thebibliography}
\end{document}